\documentclass[sigconf,authorversion,nonacm]{acmart} 

\usepackage{graphicx}
\usepackage{hyperref}
\usepackage{amsmath}
\usepackage{amsthm}
\usepackage{booktabs}
\usepackage{color}

\usepackage[ruled,vlined]{algorithm2e}
\SetAlFnt{\small}
\SetAlCapFnt{\small}
\SetAlCapNameFnt{\small}

\usepackage[capitalise]{cleveref}
\usepackage{subcaption}
\usepackage{arydshln}
\usepackage{accents}

\SetCommentSty{mycommfont}

\usepackage{csquotes}
\MakeOuterQuote{"}

\copyrightyear{2024} 
\acmYear{2024} 
\setcopyright{acmlicensed}\acmConference[ICAIF '24]{5th ACM International Conference on AI in Finance}{November 14--17, 2024}{Brooklyn, NY, USA}
\acmBooktitle{5th ACM International Conference on AI in Finance (ICAIF '24), November 14--17, 2024, Brooklyn, NY, USA}
\acmDOI{10.1145/3677052.3698607}
\acmISBN{979-8-4007-1081-0/24/11}

\DeclareMathOperator*{\argmin}{arg\,min}

\begin{document}

\title{Simulate and Optimise: A two-layer mortgage simulator for designing novel mortgage assistance products}

\author{Leo Ardon}\authornote{Shared first authorship}
\affiliation{
JP Morgan AI Research, London, \country{United Kingdom}
}

\author{Benjamin Patrick Evans}\authornotemark[1]\authornote{benjamin.x.evans@jpmorgan.com}
\affiliation{
JP Morgan AI Research, London, \country{United Kingdom}
}

\author{Deepeka Garg}
\affiliation{
JP Morgan AI Research, London, \country{United Kingdom}
}
\author{Annapoorani Lakshmi Narayanan}
\affiliation{
JP Morgan AI Research, New York, \country{USA}
}
\author{Makada Henry-Nickie}
\affiliation{
JPMorganChase Institute, Washington DC \country{USA}
}
\author{Sumitra Ganesh}
\affiliation{
JP Morgan AI Research, New York, \country{USA}
}

\def\shortauthors{}

\begin{abstract}

We develop a novel two-layer approach for optimising mortgage relief products through a simulated multi-agent mortgage environment. While the approach is generic, here the environment is calibrated to the US mortgage market based on publicly available census data and regulatory guidelines. Through the simulation layer, we assess the resilience of households to exogenous income shocks, while the optimisation layer explores strategies to improve the robustness of households to these shocks by making novel mortgage assistance products available to households. Households in the simulation are adaptive, learning to make mortgage-related decisions (such as product enrolment or strategic foreclosures) that maximize their utility, balancing their available liquidity and equity. We show how this novel two-layer simulation approach can successfully design novel mortgage assistance products to improve household resilience to exogenous shocks, and balance the costs of providing such products through post-hoc analysis. Previously, such analysis could only be conducted through expensive pilot studies involving real participants, demonstrating the benefit of the approach for designing and evaluating financial products.
\end{abstract}

\maketitle

\section{Introduction}
\label{sec.introduction}
\let\thefootnote\relax\footnotetext{\textbf{Accepted at the 5th ACM International Conference on AI in Finance 2024}} 

Designing financial products for assisting homeowners during mortgage related financial distress is a complex task due to the intricacies of the mortgage system. Traditionally, such development has relied on conducting human pilot studies to analyse the effectiveness and impact of such products \textit{in vivo}. However, such pilot studies are typically expensive and as a result, are limited to small-scale analysis, restricting the exploratory capabilities. To address these issues, a way to conduct such analysis \textit{in vitro} is needed.

This paper introduces a computational approach to optimise and evaluate novel mortgage assistance products using a two-layer reinforcement learning (RL) and simulation structure. In this approach, the outer (product) layer is responsible for optimising or selecting new product configurations, while the inner (simulation) layer employs agent-based simulation to assess the impact of these product configurations in a simulated mortgage environment. With this approach, we can conduct extensive large-scale analysis across a range of products \textit{in vitro}. The benefits of the simulation are two-fold: First, simulation enables the evaluation of a broader range of product configurations across different scenarios, at a significantly faster pace and substantially lower cost compared to traditional studies. Second, we can introduce and evaluate these products in a controlled setting, mitigating the risk of introducing potentially harmful financial products into the real market. The simulation serves as an exploratory platform before performing studies or introducing products \textit{in vivo}.\\

\noindent The specific contributions of the paper are:
\begin{itemize}
    \item A novel two-layer approach for optimising products through multi-agent simulation
    \item An extension of a mortgage servicing ABM (\cite{mortgageABM}) with a more robust representation able to perform counterfactual analysis through product conditioned policy learning
    \item A generic parameterised financial product configuration compatible with the conditional policy learning
    \item Analysis into the financial resilience of households and costs of providing cover under different product configurations 
\end{itemize}

The remainder of the paper is organised as follows. Section~\ref{secBG} provides an overview of related work and background to the area. Section~\ref{secApproach} introduces the novel two-layer approach. The impact of different mortgage assistance products is analysed in Section~\ref{secAnalysis}. Discussion and conclusions are presented in Section~\ref{secConclusion}.

\section{Background and Related Work}\label{secBG}
Several agent-based models (ABMs) have been developed for analysing housing markets \cite{geanakoplos2012getting,carro,carro2023taming,evans2021impact,merHo2023high}; however, these models have fixed behaviour rules encoded through behavioural heuristics. To model the dynamic nature of household behaviour and the autocurricula that arise from changes to market conditions, \cite{mortgageABM} developed an ABM where households learn to adapt their behaviour in response to market conditions, recreating several stylised facts of the actual mortgage market. However, the authors focused solely on a fixed mortgage product (Mortgage Reserve Accounts \cite{prosperityNow}), providing no way to generalise across products and, more importantly, no way to optimise for new products. In this work, we significantly extend upon the simulator introduced by \cite{mortgageABM}, allowing for automated product design in a two-layer framework.

Recent work has investigated the application of a two-layer framework for economic policy design \cite{zheng2022ai}, and other work treating these two-layer frameworks as Stackelberg Equilibria \cite{gerstgrasser2023oracles}, both highlighting the potential of such approaches. However, thus far, work in this space has been limited to relatively simple domains.

Despite the existence of these simulators and recent developments in automated policy design, existing mortgage assistance products tend to be manually crafted by domain experts. For example, one existing product in this domain is mortgage reserve accounts (MRA) \cite{farrell2019trading,goodman2023using}, which provide mortgage funds (for use during financial distress) without any cost to the household. An alternative formulation is matched MRA \cite{prosperityNow}, requiring $\$x$ to be provided by the household, which is then matched by the product provider. During COVID, under the CARES act, a new special COVID forbearance product was introduced \cite{corcoran2020economic}. However, thus far, these products have not been automatically optimised \textit{in vitro}, and instead deployed or explored through pilot studies \textit{in vivo}. On the contrary, our approach advances upon these ideas by integrating an adaptive outer layer with a realistic agent-based simulator to design mortgage products to improve household resilience.

To the best of our knowledge, this work proposes the first realistic census-calibrated model for product design and analysis within the US mortgage market, offering a valuable tool for policy-makers. Although our model is calibrated for the US market, the underlying approach is applicable to general housing markets.

\section{Proposed Approach}\label{secApproach}

\begin{figure}[!htb]
    \centering
    \includegraphics[width=.95\columnwidth]{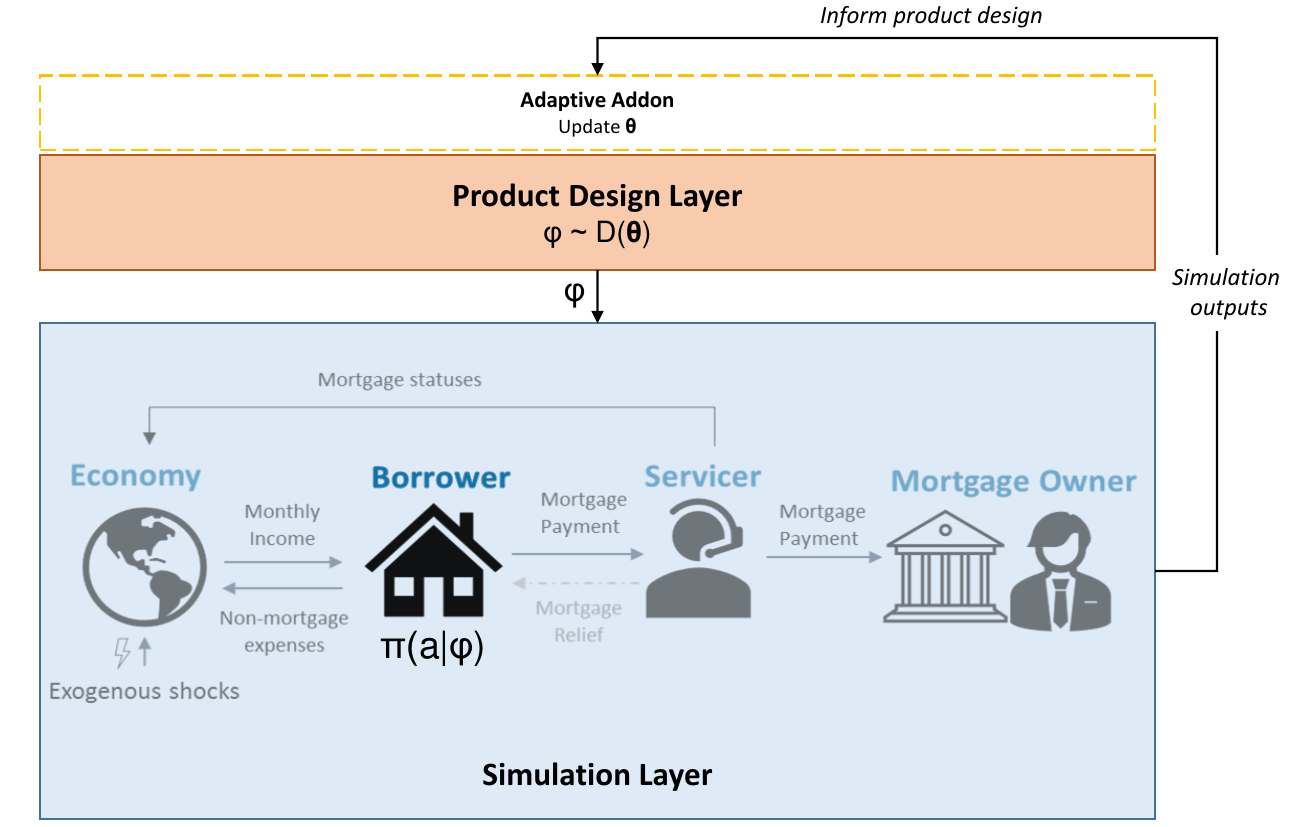}
    \caption{The proposed two-layer approach, featuring an outer (product) layer, and an inner (simulation) layer. The inner layer is conditioned on the output from the outer layer.}
    \label{figApproach}
\end{figure}

We develop a two-layer approach for mortgage product optimisation, illustrated in Fig.~\ref{figApproach}. This approach consists of a product design layer and a simulation layer. The product layer introduces novel mortgage assistance products, and the simulation layer models the impacts of these products on the mortgage ecosystem, with households dynamically adjusting their product uptake behaviour. The integration of both layers is essential for capturing the effect new products have on household behaviour and the overall mortgage ecosystem. The interaction between the two layers is illustrated in Fig.~\ref{figApproach}. The inner layer optimises the households' policy $\pi$, and the outer layer the product distribution parameters $\boldsymbol{\theta}$. %

Conducting thousands of adaptive simulations can be time consuming as agents must learn to adapt their behaviour in each scenario. Instead of requiring retraining and re-running the simulation for every product (potentially thousands of such products), we propose a method for learning generic product conditioned behaviour. This approach eliminates the need for retraining for each new product, significantly improving efficiency. In the real world, individuals can generalize and assess the benefits of products based on past experiences or similar product configurations. Similarly, we propose an approach to capture this generalization in simulation, where agents learn the impact of various product configurations through simulated experiences.

\subsection{Simulation Layer: Mortgage Ecosystem}
The simulation layer extends the model of \cite{mortgageABM}, to account for household behavior conditioned on different product configurations. Notably, while households in  \cite{mortgageABM} learned to adapt their behaviour, they lacked the ability to learn how specific products influenced their projected outcomes. This limitation stemmed from the fact that, during training, households could only be exposed to one product. As a result, learning product-specific behavior required retraining under every new scenario, significantly limiting the exploratory capabilities of their simulator. In contrast, our work introduces a novel approach for households to learn the impact of various products on their financial trajectories, enabling the analysis of novel mortgage products.

\subsubsection{Agent types}

There are four key agent types in the model: Borrowers, Servicers, Mortgage Owners, and the Economy agent. These agents are summarised below, and the overall interaction among these agents is demonstrated in Fig.~\ref{figApproach}. One timestep $t$ in the simulation corresponds to $1$ month in the real ecosystem, aligning with the standard mortgage payment schedule.

\begin{table}[]
\caption{Borrower Characteristics}\label{tblCharacteristics}
\begin{tabular}{@{}ll@{}}
\toprule
Attribute & Source                                 \\ \midrule
$I_0$    & S1901 US Census                        \\
$\$m | I_0$ & S2506 US Census                        \\
$E_0 | I_0$    & Consumer Expenditure Survey            \\
$A_0 | I_0$    & Survey of Income Program Participation \\ \bottomrule
\end{tabular}
\end{table}

\paragraph{Borrowers}\label{Borrowers} Borrowers are households that hold mortgages. Borrowers are strategic, making decisions $a_t$ each timestep that affect not only their own financial and home-ownership trajectories but can influence the overall dynamics of the broader ecosystem. Each borrower has a realistic financial profile including their income $I_t$, expenses $E_t$ (including mortgage expense $\$m$), and amount of savings $A_t$, all sampled according to census and publicly available datasets (detailed in Table~\ref{tblCharacteristics}) at $t=0$, as well as a random mortgage term remaining (from $0\dots30$ years), giving initial equity at $t=0$. 

Borrowers are adaptive, learning policies $\pi$ (via proximal policy optimisation and generalized advantage estimation \cite{schulman2017proximal}) to maximise their expected lifetime utility $U$:

\begin{equation}
\pi(a|s, \varphi) = \max_{\pi}\mathbb{E}_{\pi}\left[\sum_{t=0}^\infty\gamma^t U_t(a_t|s_t, \varphi) \right]
\end{equation}
where the borrowers' monthly utility function is a mixture of equity and liquidity:
\begin{equation}\label{eqUtility}
\begin{split}
U_t &= \gamma \cdot \text{Liquidity}_t + (1-\gamma) \cdot h_t \cdot \text{Equity}_t\\
\text{Liquidity}_t &= 1 - \min(\frac{E_t^h}{I_t}, 1)\\
\text{Equity}_t &= \frac{\text{Total Paid}}{\text{Total loan amount}} \\
\end{split}
\end{equation}

The equity component of $U_t$  encodes desire for home-ownership, and the liquidity quantifies preference for currently available cash flows,  driven by their preferences encoded via $\gamma$. Both terms have been shown to be important to mortgage behaviour \cite{farrell2019trading}. When making decisions, the borrowers observation space $s_t$ includes their preferences $\gamma$ (to allow learning a general single policy over the full spectrum of preferences), financial characteristics, and market status, including Liquidity, Equity, $h$, and their financial buffer $B_t = \frac{A_t}{E_t^h}$. Importantly, borrowers also condition their policy on the parameters of a mortgage product $\varphi$, learning product-specific behaviour for more effective financial management (through $U$). This is a crucial development, as this enables households to make utility-maximising decisions based on the specific product offered. The product details for $\varphi$ are detailed in Section~\ref{secProductLayer}.

Each timestep, a shock is (stochastically) applied to the borrower's income (more details about the shock model in the \textit{Economy} section below). The borrower agent is designed to make (up to) 2 separate decisions to accommodate for this change in their income stream. The first decision the borrower must take concerns payment (Payment Decision $\pi_1$). The borrower needs to decide whether it's in their best interest to skip their mortgage payment, pay their mortgage payment, or pay their mortgage payment and opt-in for any available mortgage product. The second decision (Relief Decision $\pi_2$) is only available following any missed payments by the borrower, and is related to the relief options offered by the servicer (detailed in the \textit{Servicer} section below), to help the borrower navigate financial hardships. Upon missing a payment, servicers can offer relief to the borrower, and borrowers can decide if they want to accept or reject the relief offer provided by the servicer. For notational clarity, we refer to $\pi = \{\pi_1, \pi_2\}$.

To permit efficient computation, borrowers learn a shared (but individualised) policy to maximize their $U$, using the concept of agent supertypes \cite{vadori2020calibration}, where the policy is conditioned on agent specific characteristics, allowing for heterogeneity in the resulting behaviour \cite{evans2024learning} but more scalable training. For example, borrowers with $B_t \to 0$ may see more benefit from the cover provided by a mortgage product and decide to enrol, whereas borrowers with high financial buffers can withstand shocks and may decide not to enrol to avoid paying the product fees. 

\paragraph{Servicers} Mortgage servicers are responsible for managing all month-to-month loan activities, serving as the intermediary between borrowers and mortgage owners. As we are focused on the US market, during times of financial distress, servicers are required to offer borrowers loan support options (outlined in Table~\ref{tblServicer}), in accordance with the regulatory framework "Regulation X"\footnote{12 CFR Part 1024 - Real Estate Settlement Procedures Act}, ensuring regulatory-guided servicing practises. That is, before foreclosure, servicers must offer distressed borrowers alternative options, including organising repayment plays, providing temporary forbearance, or reducing the monthly loan payments expected from the borrower. It is up to the borrower whether they chose to accept or reject these relief options. The environment is generic enough to allow alternative relief options for other countries, where regulatory policies may differ.

\begin{table}
\caption{Servicer relief hierarchy, based on  Regulation X (simplified) from the Consumer Financial Protection Bureau.}\label{tblServicer}
\centering
\resizebox{\columnwidth}{!}{%
\begin{tabular}{@{}lll@{}}
\toprule
Order & Relief   Type     & Adjusted   Loan Payments                                                                                                                                                                        \\ \midrule
1     & Repayment Plan    &  Pay missed + Resume \\
2     & Forbearance       &  3 Month pause                                                                                \\
3     & Loan Modification &  20\% reduction                                                                         \\
4     & Foreclosure       &  None (mortgage ended, property sold)                                                            \\ \bottomrule
\end{tabular}%
}
\end{table}

\paragraph{Mortgage Owners} Mortgage owners can be financial institutions (bank or non-bank), investors, or government-sponsored enterprises (GSE), that own the underlying mortgage. 

\paragraph{Economy} The dynamic nature of the wider economy is modeled with an economy "agent". The economy agent controls the economic environment of the simulation at $t$ by:
\begin{itemize}
    \item applying shocks to the borrower’s monthly income $I_t$
    \item updating the House Price Index $h_t$ as function of the number of foreclosures observed (\cref{eqForeclose})
\end{itemize}
	
Income $I_t$ is modelled as a random walk subject to exogenous shocks to simulate various life events one may experience (e.g., job loss, pay raise etc.) that can impact borrowers' monthly income either positively or negatively. Shocks have a size 
$s$ from $S = [-1, -0.9, \dots, -0.1, 0, +0.1, \dots, +0.9, +1]$. Shocks effect monthly income as $$I_t = I_{t-1}*(1+s)$$ for example, $s=0$ does not impact income, $s=-1$ entirely wipes out income, and $s=1$ doubles income. Intermediary $-1<s<1$ proportionately adjusts borrower income. During training, a shock has a likelihood of $\frac{1}{12}$ of occurring (averaging one shock per year), with uniform likelihood among $S$ (as a result, a lower probability of experiencing a shock with magnitude larger than $x$ as $x$ increases). During evaluation, fixed size shocks $s$ are applied at a given timestep $t_s$. 
The impact of the decisions of other borrowers on mortgage decisions is encoded indirectly through the market status $h_t$. Intuitively, a higher number of foreclosures drives house prices down, as sellers (e.g. servicers) tend to sell the property below its market value to close as quickly as possible to recoup their fees and limit \begin{equation}
h_t=h_{t-1}\cdot(1-\#\text{foreclosures}_{t-1} \cdot \delta)\label{eqForeclose}
\end{equation}
where $\delta$ is a hyperparameter representing the effect that one foreclosure has on the HPI, creating a feedback loop and interference among the learning agents.

Additional extensions to the economy could consider the impact of other exogenous macroeconomic factors, such as interest rates, or inflation.

\subsection{Optimisation layer: Product Design}\label{secProductLayer}
We are interested in designing financial assistance products for borrowers. Successful products should meet several desiderata. The products should be \textit{D1}.) Simple enough to understand, \textit{D2}.) Relate to the existing available products, and \textit{D3}.) Flexible to allow for novel extensions.


\subsubsection{Parameterised Product}
To achieve these desiderata, we propose a generic \textit{parameterised} product to generalise across the product space. The product is represented as:

\begin{equation}\label{eqBasicFormat}
    \varphi( P_0, P_t, V, \dots)
\end{equation}
covering three key parameters: the upfront premium  $P_0 \geq 0$ (e.g. enrolment cost), the ongoing monthly fee $P_t \geq 0$, and the total cover value provided by the product $V \geq 0$. These three parameters align with well-established actuarial terms \cite{letters1949international}, however, the product formulation remains general enough to accept product-specific parameters.

For borrowers that choose to opt into a mortgage product, their monthly housing expenses are adjusted to reflect the product costs and cover provided $E_t^h = m + P_t - \chi_t(V)$, where $\chi_t$ returns the amount of cover used at time-step $t$.



\paragraph{Existing products as special cases} As introduced in Section~\ref{secBG}, the MRA product can be represented with the following parameterization: 

\begin{equation}
    \varphi_\text{MRA}(P_0 = 0, P_t=0, V=\$x)
\end{equation}
where $\$x$ is the reserve amount, provided as product cover, at no upfront or recurring cost to the household.

Likewise, matched MRA can be encoded by:
\begin{equation}
    \varphi_\text{Matched MRA}(P_0 = \$x, P_t=0, V=\$2x)
\end{equation}
requiring an upfront enrolment cost of $\$x$, but receiving $\$2x$ in cover. 

Other products, such as COVID forbearance can also be regarded as a specific instance of the proposed product:

\begin{equation}
    \varphi_\text{COVID}(P_0 = 0, P_t=0, V=0, F \in \{3,6, 12\})
\end{equation}

Where $F$ corresponds to the length (in months) of the forbearance. This formulation leverage the ability for our product formulation to accept product-specific parameters. In practice the ability to provide forgiveness $F$ typically can only be provided at the governmental level (as was done during COVID), due to the multiple parties involved in a mortgage.

\paragraph{General case}
These aforementioned products are considered unsustainable due to the challenge of funding the cover, prompting investigation of alternative mutually beneficial (for both the household and product provider) products. Consequently, in this work, we consider monthly premium and cover conditioned on the monthly mortgage payment $\$m$:

\begin{equation}
    \varphi_*(P_0= p_0 \times \$ m, P_t= p \times \$m, V=v \times \$m)
\end{equation}
where $0 \leq p_0, p \leq 1$ adjust the pricing of the enrolment and going cost, which is upper bound at $1$ to ensure the fees never exceed the base mortgage payment amount.  $0 \leq v \leq T$ adjusts the value of the cover, upper bound to be full cover for the duration of the simulation episode (we assume no cover replenishment here). The parameterised products have three variables, where we categorise the range of these variables using a trivariate distribution $D$.

\subsubsection{Outer layer configuration}
We propose two possible setups for the product design layer, a fixed layer, and an adaptive layer. Each setup involves distinct methods for sampling and presenting new products to the households through $D$.

\textbf{Fixed layer} The fixed layer samples products from fixed distributions, e.g. $\varphi \sim D$. For this work, the fixed layer samples from a multivariate uniform distribution  $D = \{\mathcal{U}(0, 1), \mathcal{U}(0, 1), \mathcal{U}(0, T)\}$ for simplicity, but any distribution could be used.

\textbf{Adaptive Layer} The adaptive layer takes this one step further by also adapting the product distribution, learning suitable parameters for the distributions to minimise some loss function $\mathcal{L}$, i.e.  $\boldsymbol{\theta} = \argmin_{\theta} \mathcal{L}$. As an example, the adaptive layer may learn more promising regions are instead $\{\mathcal{U}(0.5, 0.7), \mathcal{U}(0.25, 0.3), \mathcal{U}(9, 12)\}$, focusing on one particular region of the space through updating $\boldsymbol{\theta}$, generating more specialised products. While the representation allows for any optimization algorithm to output an updated $\boldsymbol{\theta}$, here, we use a RL-based approach, taking actions 
$$\mathbf{a} = \{\Delta \mu_{p_0}, \Delta \delta_{p_0}, \Delta \mu_{p_t}, \Delta \delta_{p_t}, \Delta \mu_{V}, \Delta \delta_{V} \}$$
that adjust the boundaries of the underlying multi-variate distribution, where the lower bound is $\mu - \delta$, and upper bound is $\mu + \delta$, using the optimization metric $-\mathcal{L}$ as reward signal driving the update. The motivation for learning $\Delta$'s rather than direct parameters is to incrementally adjust the paramaters, approximating a two-time scale algorithm \cite{fiez2019convergence}, with the adaptive layer operating at a slower timescale to the simulation layer, giving sufficient time for the simulation layer to adapt to the changes. Learning the $\Delta$'s helps to smooth this bi-level updating process. Pseudo-code for this process is described in Algorithm~\ref{algAdaptive}.

\begin{algorithm}[!t]
\SetAlgoLined
\KwIn{
$\boldsymbol{\theta}=$ Fixed product parameter distribution \\
\\
\Indp \Indp
$\boldsymbol{\pi}=$ Random policy}
\KwOut{\\
\Indp \Indp
$\boldsymbol{\theta}=$ Optimized product parameter distribution\\
$\boldsymbol{\pi}=$ Trained policy
}
 
\For{each iteration}{

    \tcp{Inner layer}
    traces $\gets \varnothing$\;
    
    \For{each simulation step}{
        traces $\gets$ Simulation($\boldsymbol{\pi}, \boldsymbol{\theta})$\;
    }
    
    $\boldsymbol{\pi} \gets $RL($\boldsymbol{\pi}$, traces)\;
    \BlankLine

    \tcp{Outer layer}
    
    loss  $\gets \mathcal{L}(f(\text{traces}))$\;
    
    $\boldsymbol{\theta}  \gets$ Optimization($\boldsymbol{\theta}$, loss)\;

 }
\caption{Two-layer adaptive process}\label{algAdaptive}
\end{algorithm}

The decision to use a fixed or adaptive outer layer can be made based on whether the product designer, policy maker, or researcher has \textit{a priori} known preferences for the product outcomes (encoded in some loss function $\mathcal{L}$). In such cases, the adaptive configuration should be chosen, as it allows narrowing the search space to the regions optimizing $\mathcal{L}$. If instead, one only wants to make the decision \textit{post-hoc} after considering all product configurations, then the fixed outer layer can be used, as this option provides a range of potential product configurations for evaluation. Likewise, to evaluate a specific product, a fixed layer can be used treating $D$ as a multivariate dirac delta, with probability mass concentrated solely at the product of interest, or by specifying a fixed range for evaluating specific products. In this work, we provide the ability to employ both fixed and adaptive approaches, demonstrating the flexibility of the proposed approach through specifying $D$.

\section{Experimental Results and Analysis}\label{secAnalysis}

The experiment details and results are outlined in this section. The goal is to see if through the two layer representation, we can effectively develop novel mortgage assistance products to improve the financial resilience and stability of borrowers in the mortgage ecosystem.

\subsection{Configuration}

\subsubsection{Metrics of Interest}\label{secMetrics}


\paragraph{Delinquency rate} A borrower is considered \textit{delinquent} when they miss a mortgage payment, e.g. mortgage payment made $<\$m$. The delinquency rate $0 \leq r \leq 1$ is the proportion of delinquent borrowers. If a product is useful to borrowers, $r$ will reduce, showing that borrowers' mortgage payments were able to withstand shocks. Conversely, if a product is harmful and misinformed borrowers opt for it, $r$ will increase (e.g. if the product fees were more costly than the cover used). In cases where a product is harmful, but borrowers are strategic enough, they will learn to not take the product, resulting in an unchanged $r$. 

\paragraph{Social index} To gain a more nuanced understanding of the delinquency rate across subgroups of the population, we utilise a social index $-\omega$ based on the worst delinquency rate across the borrower groups. The goal is to reduce the delinquencies, increasing social welfare in the system. The following definition is used:

\begin{equation}\label{eqRule}
   \omega(r) = \max_{g \in G} r_g
\end{equation}
i.e., the min-max rule from social choice theory \cite{sen2018collective}, stating that we want to minimise $\omega(r)$, and thus improve the most affected group by reducing the proportion of delinquent borrowers, improving the worst-case across the population. To group borrowers, we use a grouping criteria based on the Mission Index Criteria (MIC) developed by government-sponsored enterprises \cite{MissionIndexFramework}. The MIC aims to identify borrower groups needing additional support \cite{sustainalytics.com_2024}. Specifically, the key selection criteria for borrowers to be part of the index is based on the median income, with eligible low-income borrowers defined as those with an annual income $<80\%$ of the Area Median Income (AMI)\footnote{We do not analyze the geographic element in this work, so consider the US-wide area.}. Although additional factors (such as census tracts) influence the selection criteria, in this work, we focus on the low-income threshold. Consequently, we separate borrowers into two groups: $G=\{$low income, other$\}$, based on this 80\% threshold.


\paragraph{Product Cost} The product cost determines what it costs per borrower for the product provider to offer the product. In order to calculate the effective product cost for an individual borrower, the following definition is used:
\begin{equation}
    C_n = \chi(V) - P_0 -\sum_{t=i_n}^T P_t 
\end{equation}
where $i_n$ is the enrolment timestep ($i_n=\infty$ if never enrolled), and $\chi(V)$ is a function returning the amount of cover actually used by borrower $n$ throughout the episode ($\chi(V) = \sum_{t=0}^T \chi_t(V)$). The aggregate product cost is determined as:
\begin{equation}
    C = \frac{1}{N} \sum_{n =0}^N C_n
\end{equation}
which is the general cost of offering the product across borrowers. Product providers want to minimise the aggregate cost $C$, and borrowers indirectly wish to minimise their individual cost $C_n$ through the optimisation of their utility (\cref{eqUtility}).




\subsubsection{Comparison}

As a baseline, we compare against the case of \textbf{no product} $\varphi_\varnothing$. This can be represented as a dummy product with no cost or cover:

\begin{equation}
    \varphi_\varnothing = \varphi ( P_0=0, P_t=0, V=0)
\end{equation}

Additionally, as mentioned in Section~\ref{secProductLayer}, existing products such as MRA can be seen as a special case of the proposed product, contained fully within the parameterisation proposed, so the comparison is implicit through the sampling scheme. Based on the results, we can tell that if MRA arises, then MRA is a beneficial product.

\subsubsection{Calculation}

The evaluation is conducted based on the metrics detailed in Section~\ref{secMetrics} across exogenous negative income shocks:  $S^- = \{s \in S | s \leq 0 \}$.  To compare metrics across $S^{-}$, we compute the integral under the shock curve (using the trapezoidal rule), converting the shock-specific metric into a single scalar for comparison purposes. Unless a metric is explicitly conditioned on $s$, it should be assumed to be the integral across $S^{-}$.

\subsection{Baseline Results}

\begin{table}
\center
\caption{Metrics of interest $F$ integrated across shocks $S^-$. For each setting, the best product is displayed. For the outer layer, additionally the average product (mean $\pm$ standard deviation) sampled from $\boldsymbol{\theta}$ is displayed (this is not included for the baseline as there is only a single dummy product). Lower values are better, and the best approach for each row is \textbf{bolded}.}\label{tblMetrics}
\centering
\resizebox{.8\columnwidth}{!}{
\begin{tabular}{@{}lccc@{}}
\toprule
 & & \multicolumn{2}{c}{\textbf{Two Layer}} \\ 
 $F$ & Baseline & Fixed & Adaptive \\
 \midrule
$\boldsymbol{\omega(r)}$ &  &  &  \\
- Best product & 0.536 & 0.03  & \textbf{$<$ 0.01} \\
- Mean of region &  \textit{NA} & 0.46 $\pm$ 0.12 & \textbf{0.25 $\pm$ 0.2} \\
 &  &  &   \\  \hdashline
 &  &  &   \\
$\mathbf{C}$ &  &  &   \\
- Best product & 0 & \textbf{-7164} & -302  \\
- Mean of region & \textit{NA} & \textbf{1052 $\pm$ 3696}   & 7848 $\pm$ 7951 \\ \bottomrule
\end{tabular}
}
\end{table}

\begin{figure}[!htb]
\centering\includegraphics[width=.9\columnwidth]{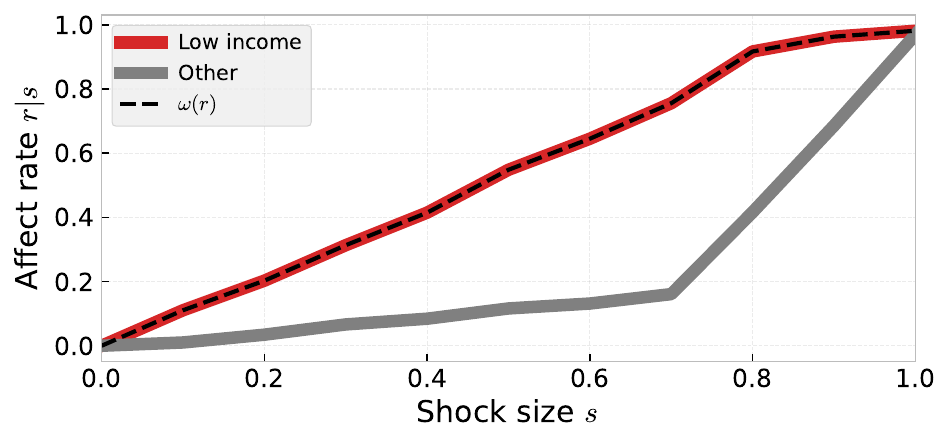}
\caption{The baseline delinquency rate $r$ and (negative) social index $\omega(r)$ under $\varphi_\varnothing$.}
\label{figBaseline}
\end{figure}

The delinquency rate $r$ and (negative) social index $\omega(r)$ 
conditioned on $s$ are illustrated in Fig.~\ref{figBaseline} for the baseline setting $\varphi_\varnothing$. The resulting (integrated) metrics across $S^{-}$ are given in Table~\ref{tblMetrics}. The dummy product $\varphi_\varnothing$ does not incur a product cost, i.e. $C = C_n = 0$, so we have ommited these plots.

Fig.~\ref{figBaseline} helps to understand the stability and robustness of the two groups under negative income shocks. It is evident that the lower income group is significantly less robust to these relative shocks, having higher delinquencies, thus necessitating an increased focus on improving the mortgage stability of these borrowers, e.g., to improve investment and lending conditions, through more sustainable home-ownership. The long term sustainability for these low-income borrowers is of particular concern (only strengthened following the 2008 crisis) \cite{van2011sustainability}, therefore, developing products that help improve the robustness of borrowers among these groups is essential to foster home-ownership opportunities. In the next section, we therefore attempt to develop new products to improve this robustness (by reducing the delinquencies).

\subsection{Product Evaluation}
To find what products best improve outcomes for these groups, products are evaluated across the two metrics $F=\{\omega(r)$, $C\}$. These metrics are conflicting, as it is costly to provide the cover needed to optimise $\omega(r)$. 

\subsubsection{Fixed product layer}

\begin{figure}[!htb]
    \centering
    \begin{subfigure}{.48\columnwidth}
      \centering
      \includegraphics[width=\textwidth]{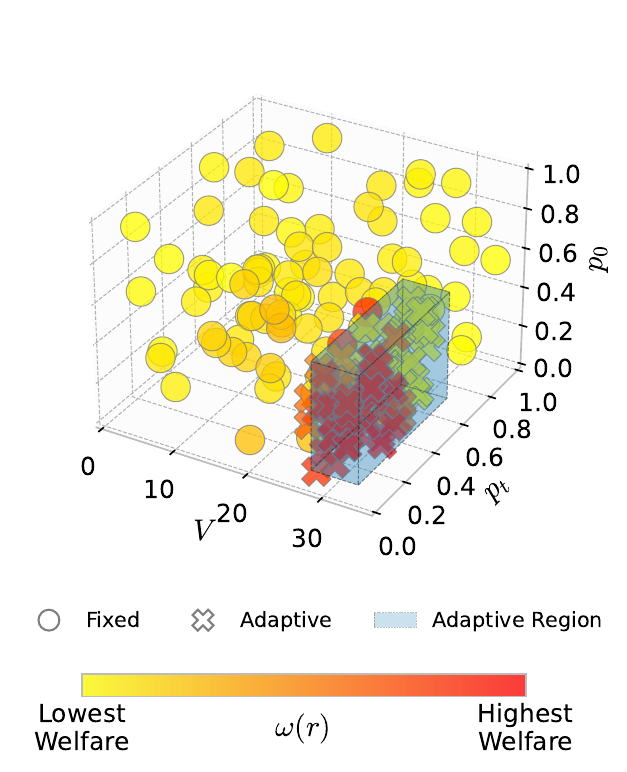}
      \caption{Delinquency rate $\omega(r)$}
      \label{figMetricRate}
    \end{subfigure}
    \begin{subfigure}{.48\columnwidth}
        \includegraphics[width=\textwidth]{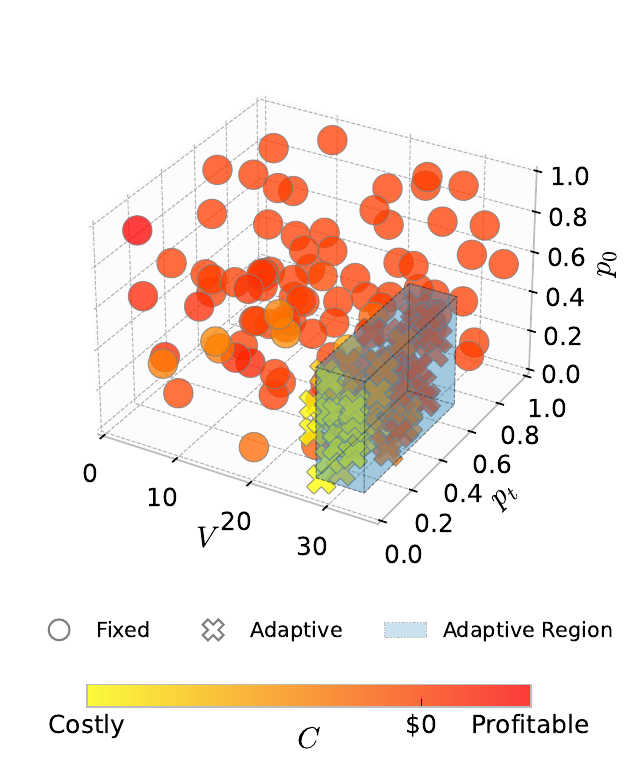}
        \caption{Aggregate cost $C$}
        \label{figMetricC}
    \end{subfigure}%
\caption{Product analysis across configurations. Products sampled from the fixed (adaptive) outer layer are shown with $\circ$ ($\times$), with $100$ samples each. The region identified by the adaptive layer is indicated by the blue bounding box. In all cases darker colour indicates a \textbf{better} value.}
\label{figMetricAffects}
\end{figure}

For the fixed product layer, we consider a multivariate uniform distribution across the product parameter space (from Section~\ref{secProductLayer}), sampling randomly from this space during training. A visualisation of how different sampled product configurations affect the metrics is given in Fig.~\ref{figMetricAffects}, providing a clear overview of what regions of product configurations are beneficial for each metric. For example, for $\omega(r)$ (Fig.~\ref{figMetricRate}), the combination of high cover, low enrolment cost, and low monthly fees provides the best products. However, viewing Fig.\ref{figMetricC} we can see that these products generally come at the highest cost for the provider. Importantly however, profitable products for the provider are not only those with the lowest cover and higher costs, as these may not see uptake in the simulation, instead a balance is preferred, and providers can still see good profit from higher cover products, demonstrating one benefit of the simulation for modelling product uptake. The integrated metrics are presented in Table~\ref{tblMetrics}, demonstrating the outer layer successfully finds products which improve over the baseline under both metrics. 


\subsubsection{Adaptive product layer}

The results above assume there is no \textit{a priori} known trade-off among the metrics $F$. If there is a preference among these objectives, e.g.:
$$\mathcal{L}: \mathbb {P} (F) \to \mathbb{R}$$
where  $\mathbb {P} (F)$ is the powerset of $F$,
we can instead use an adaptive product layer with $\mathcal{L}$ as our optimisation criteria, finding product parameter regions most compatible with the desired outcome. 
To demonstrate that the adaptive product layer works as intended (finding a suitable $\boldsymbol{\theta}$), we evaluate across the social index, showing that a logical region is found that optimises $\mathcal{L}$.

\paragraph{$\mathcal{L}(\omega(r))$} Maximising the social index (minimising $\omega(r)$)  results in the product region of interest as visualised by the bounding box in Fig.~\ref{figMetricRate}. Comparing to the fixed layer in Fig.~\ref{figMetricRate}, a logical region has been found, with $D(\boldsymbol{\theta})$ resulting in products that successfully minimise the overall delinquency rate, increasing the social index, and improving upon the rates achieved by the fixed layer alone (confirmed in Table~\ref{tblMetrics}). These results highlight the usefulness of the adaptive layer when the optimisation criteria is known, providing more products within desirable product regions. However, generally speaking, this improvement does come at the expense of $C$, resulting in more costly products (Fig.~\ref{figMetricC}). In the following subsection, we analyse this in more detail.


\subsection{Non-dominated products}

\begin{figure}[!htb]
    \centering
    \includegraphics[width=\columnwidth]{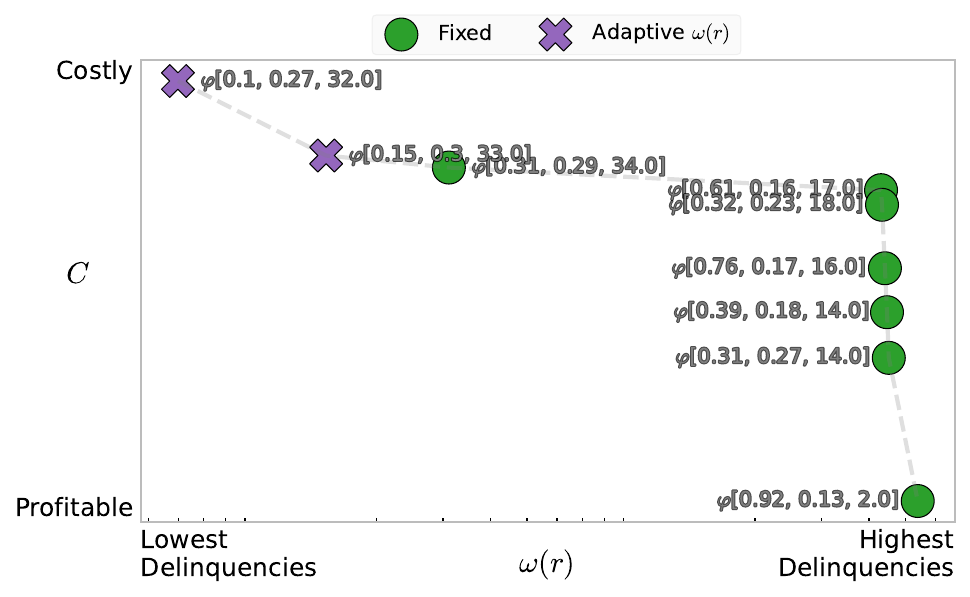}
    \caption{Pareto frontier of product configurations across the two metrics $C$, $\omega(r)$ for the products in Fig.~\ref{figMetricAffects}. The social index (x-axis) is visualised on a log-scale. The (unattainable) ideal product is situated in the bottom left corner.}
    \label{figPareto}
\end{figure}

To understand the implications for the products generated through the optimization of the social index, in this section we evaluate the tradeoffs among the metrics. To do this, we consider Pareto dominance. A product Pareto dominates another product if:

\begin{equation} \label{eqDominance}
\varphi^{i} \succ_{P} \varphi^j \Leftrightarrow \big( \forall f \in F; \; f_{\varphi^{i}} \leq f_{\varphi^{j}} \big) \wedge \big( \exists f \in F; \; f_{\varphi^{i}} < f_{\varphi^{j}} \big)
\end{equation}
e.g. product $\varphi^{i}$ is at least as good as product $\varphi^{j}$ on all metrics, and strictly better on at least one\footnote{Note that this definition of $\succ_{P}$ assumes we are minimising the metrics, but similar definition holds when maximising (with $-f$)}. The Pareto frontier is the set of non-dominated products, where there is no other (considered) product that can improve on any metric without worsening another, providing an ideal set of products for consideration when no other information is known. 

We visualise the resulting Pareto frontier across the fixed and adaptive layers (from the products in Fig.~\ref{figMetricAffects}) in Fig.~\ref{figPareto}, demonstrating that the adaptive layer is able to find more specialised products of interest than the fixed layer setting, providing additional products in the region we are focused on (two new product configurations improving upon the maximum social index - or reducing delinquencies - from the fixed layer setting), whereas the fixed layer has products distributed more broadly across the product space (e.g. more profitable products at the expense of $\omega(r)$). This frontier shows the benefit of the adaptive layer -- generating more products in the region of interest, and the benefit of the fixed layer -- generating a wider range of products when preferences are unknown \textit{a priori}.  The non convexity of the frontier is due to the non-linear product uptake behaviour of the agents, helping to show the benefit of simulating such uptake beyond a trivial assumption of convexity of uptake across the product space.

These frontiers are useful for policy-makers or institutions to analyse the trade-offs of these products, choosing to offer the products or complete further pilot studies based on these results. For example, if mortgage servicers are providing the product, this trade-off may be analysed by comparing the product cost to their internal cost of handling mortgage defaults. These products allow servicers to consider alternative approaches to prevent these costly and often time-consuming default handling operations, and potentially help with the alignment of servicer incentives \cite{diop2023mortgage}. Non-bank servicers are particularly sensitive to cash flow and liquidity risks \cite{kaul2020improving,harrelson2021industry} so consideration of such alternatives is important for managing these risks. Interestingly, existing MRA-styled products do not show up on the frontier, indicating that more useful product combinations are possible, such as those identified in Fig.~\ref{figPareto}, providing important potential products for further consideration.

\subsection{Results summary}
We first verified the social index following income shocks under the baseline setting, before then expanding to the integration of novel products to improve upon this social index. Under this setting, we evaluated our two-layer approach. We demonstrated how the outer layer can be used for analysis across a wide range of products, either through a fixed or adaptive setting, successfully improving the social index across the population. Under the fixed setting, no \textit{a priori} preference towards objectives is required, and instead, such decisions can be made post hoc by analysing the resulting frontier. However, when information about the desired product outcome is known, the adaptive layer can dynamically find product configurations and regions of interest for best optimising these outcomes, narrowing and fine-tuning the space of products under consideration. 

These results demonstrate the suitability and flexibility of the proposed approach for product optimisation, and how such a framework can be used for analysing complex scenarios such as the maximisation of a social index in the mortgage ecosystem. While here we validated with optimising for a single metric, this generally came at the expense of worsening other metrics, as identified by the frontier analysis. Future work could consider combinations of metrics, e.g. $\mathcal{L} \bigl(C, \omega(r) \bigr) = \frac{C}{\omega(r)}$, to balance the cost of providing the product with the resulting social index, or consider multiobjective optimisation methods \cite{hayes2022practical} for simulatenously improving both metrics, but this is beyond the scope of the current analysis.


\section{Conclusion and Discussion}\label{secConclusion}

We developed a novel two-layer approach for automated financial product design, and demonstrated the use of this approach for designing mortgage products in a simulated mortgage ecosystem. Through optimising for a social index, we showed how we can improve the financial stability of households (by reducing delinquencies), which can lead to improved homeownership rates for even the most impacted groups of the population.

The proposed approach simulates product uptake and product impact within a census calibrated mortgage ecosystem, providing insights into the effectiveness of various product configurations for the US market. To analyze numerous product configurations efficiently, we introduced a generic parameterised product, and enabled borrowers to learn product-specific behaviour through conditioning decisions on these parameterizations. Borrowers learn decisions to best maximise their lifetime utility (balancing their equity and liquidity) through a policy gradient approach, collecting trajectories in the simulated mortgage environment based on different products, learning to generalize across product configurations with conditional policy learning.

The integration of this product conditioned policy learning combined with an outer parameterised product layer allowed for wide-scale analysis, a capability which was not possible with the existing simulation methods and instead, required retraining for every product, or expensive \textit{in vivo} pilot studies. We hope that the proposed approach can assist policy-makers and product designers in developing the next generation of mortgage assistance products to best serve households through extensive \textit{in vitro} analysis before deploying real pilot studies.

\balance
\section*{Disclaimer}
This paper was prepared for informational purposes by the Artificial Intelligence Research group of JPMorgan Chase \& Co and its affiliates (“J.P. Morgan”) and is not a product of the Research Department of J.P. Morgan.  J.P. Morgan makes no representation and warranty whatsoever and disclaims all liability, for the completeness, accuracy or reliability of the information contained herein.  This document is not intended as investment research or investment advice, or a recommendation, offer or solicitation for the purchase or sale of any security, financial instrument, financial product or service, or to be used in any way for evaluating the merits of participating in any transaction, and shall not constitute a solicitation under any jurisdiction or to any person, if such solicitation under such jurisdiction or to such person would be unlawful. © 2024 JPMorgan Chase \& Co. All rights reserved.

\bibliographystyle{ACM-Reference-Format}
\bibliography{bib}

\appendix
\section{Model Parameters}

Key parameters for the model are highlighted in Table~\ref{tblParams}. 

\begin{table}[!htb]
\caption{Key model paramaters}\label{tblParams}
\begin{tabular}{@{}ll@{}}
\toprule
\textbf{Paramater} & \textbf{Value} \\ \midrule
Training Iterations & 500 \\
Rollouts & 10 \\
Number households & 100 \\
Minimum Income & 12000/year \\
Max mortgage length & 30 Years \\
Foreclosure impact $\delta$ & 0.01 \\
Liquidity preference $\gamma$ & Training: [0..1], Evaluation: 0.5\\
Shock Probability & \begin{tabular}[c]{@{}l@{}}Training: 1/12\\ Evaluation: Specific times\end{tabular} \\
Shock range & \begin{tabular}[c]{@{}l@{}}Training: [0.. 2]\\ Rollout: Specific size\end{tabular} \\ \bottomrule
\end{tabular}
\end{table}

\end{document}